\documentclass{elsart}

\usepackage[square,comma]{natbib}
\usepackage{graphicx}
\usepackage{pxfonts}
\usepackage{lineno}

\usepackage{amssymb}
\usepackage{footnote}
\usepackage{multirow}
\usepackage{varwidth}
\usepackage{array}

\journal{}

\begin{document}

\thispagestyle{empty}
\begin{Large}
\textbf{DEUTSCHES ELEKTRONEN-SYNCHROTRON}

\textbf{\large{Ein Forschungszentrum der Helmholtz-Gemeinschaft}\\}
\end{Large}

DESY 13-138

August 2013

\begin{eqnarray}
\nonumber
\end{eqnarray}
\begin{center}
\begin{Large}
\textbf{Proposal to generate 10 TW level femtosecond x-ray pulses
from a baseline undulator in conventional SASE regime at the
European XFEL}
\end{Large}
\begin{eqnarray}
\nonumber &&\cr \nonumber
\end{eqnarray}

\begin{large}
Svitozar Serkez$^a$, Vitali Kocharyan$^a$, Evgeni Saldin$^a$, Igor
Zagorodnov$^a$, Gianluca Geloni$^b$
\end{large}

\textsl{\\$^a$Deutsches Elektronen-Synchrotron DESY, Hamburg}
\begin{large}

\end{large}
\textsl{\\$^b$European XFEL GmbH, Hamburg}
\begin{large}

\end{large}

\begin{eqnarray}
\nonumber
\end{eqnarray}
\begin{eqnarray}
\nonumber
\end{eqnarray}
ISSN 0418-9833
\begin{eqnarray}
\nonumber
\end{eqnarray}
\begin{large}
\textbf{NOTKESTRASSE 85 - 22607 HAMBURG}
\end{large}
\end{center}
\clearpage
\newpage

\begin{frontmatter}



\title{Proposal to generate 10 TW level femtosecond x-ray pulses from a baseline undulator in conventional SASE regime at the European XFEL}


\author[DESY]{Svitozar Serkez \thanksref{corr},}
\thanks[corr]{Corresponding Author. E-mail address: svitozar.serkez@desy.de}
\author[DESY]{Vitali Kocharyan,}
\author[DESY]{Evgeni Saldin,}
\author[DESY]{Igor Zagorodnov,}
\author[XFEL]{Gianluca Geloni,}

\address[DESY]{Deutsches Elektronen-Synchrotron (DESY), Hamburg, Germany}
\address[XFEL]{European XFEL GmbH, Hamburg, Germany}

\begin{abstract}
Output characteristics of the European XFEL have been previously
studied assuming an operation point at 5 kA peak current. In this
paper we explore the possibility to go well beyond such nominal peak
current level. In order to illustrate the potential of the European
XFEL accelerator complex we consider a bunch with 0.25 nC charge,
compressed up to a peak current of 45 kA. An advantage of operating
at such high peak current is the increase of the x-ray output peak
power without any modification to the baseline design. Based on
start-to-end simulations, we demonstrate that such high peak
current, combined with undulator tapering, allows one to achieve up
to a 100-fold increase in a peak power in the conventional SASE
regime, compared to the nominal mode of operation. In particular, we
find that 10 TW-power level, femtosecond x-ray pulses can be
generated in the photon energy range between 3 keV and 5 keV, which
is optimal for single biomolecule imaging. Our simulations are based
on the exploitation of all the 21 cells foreseen for the SASE3
undulator beamline, and indicate that one can achieve diffraction to
the desired resolution with 15 mJ (corresponding to about $3\cdot
10^{13}$ photons) in pulses of about 3 fs, in the case of a 100 nm
focus at the photon energy of 3.5 keV.
\end{abstract}

%
%
%
\end{frontmatter}



\section{\label{sec:intro} Introduction}

Imaging of single molecules at atomic resolution using radiation
from the European XFEL facility would enable a significant advance
in structural biology, because it would provide  means to obtain
structural information of large macromolecular assemblies that
cannot crystallize, for example membrane proteins. The imaging
method "diffraction before destruction" \cite{HAJD}-\cite{SEIB}
requires pulses containing enough photons to produce measurable
diffraction patterns, and short enough to outrun radiation damage.
The highest signals are achieved at the longest wavelength that
supports a given resolution, which should be better than 0.3 nm.
These considerations suggest that the ideal energy range for single
biomolecule imaging spans between 3 keV and 5 keV \cite{BERG}. The
key metric for optimizing a photon source for single biomolecule
imaging is the peak power. Ideally, the peak power should be of the
order of 10 TW \cite{SELF0}.

The baseline SASE undulator sources at the European XFEL will
saturate at about 50 GW \cite{TSCH}. While this limit is very far
from the 10 TW-level required for imaging single biomolecules, a
proposal exists to improve the output power at the European XFEL by
combining self-seeding \cite{SELF}-\cite{ASYM}, emittance spoiler
foil \cite{EMM1}-\cite{DING}, and undulator tapering techniques
\cite{TAP1}-\cite{LAST}. However, the realization of such proposal
requires installing additional hardware in the undulator system and
in the bunch compressor \cite{SELF0}. Here we explore a simpler
method to reach practically the same result without additional
hardware. This solution is based on the advantages of the European
XFEL accelerator complex, which allows one to go well beyond the
nominal 5 kA peak current.

The generation of x-ray SASE pulses at the European XFEL using
strongly compressed electron bunches has many advantages, primarily
because of the very high peak power, and very short pulse duration
that can be achieved in this way \cite{DOHL}. Considering the
baseline configuration of the European XFEL \cite{TSCH}, and based
on start-to-end simulations, we demonstrate here that it is possible
to achieve a 100-fold increase in peak power by strongly compressing
electron bunches with nominal charge. In this way we show (see
Section \ref{sec:FELst} for more details) that 10 TW power level, 3
fs-long pulses at photon energies around 4 keV can be achieved in
the SASE regime. This example illustrates the potential for
improving the performance of the European XFEL without additional
hardware.

The solution to generate 10 TW power level proposed in this article
is not without complexities. The price for using a very high
peak-current is a large energy chirp within the electron bunch,
yielding in its turn a large (about $1 \%$) SASE radiation
bandwidth. However,  there are very important applications like
bio-imaging, where such extra-pink x-ray beam is sufficiently
monochromatic to be used as a source for experiments without further
monochromatization.

The European XFEL features three x-ray sources. Three long
undulators, operated in the SASE mode, will provide x-ray radiation
in the photon energies between 0.26 keV and 25 keV. In order to
enable high focus efficiency with commercially available mirrors (80
cm-long) at photon energies around 4 keV, the undulator source needs
to be located as close as possible to the bio-imaging instrument.
With this in mind we performed simulations for the baseline SASE3
undulator at a nominal electron beam energy of 17.5 GeV. We
optimized our setup based on start-to-end simulations for an
electron beam with 0.25 nC charge, compressed up to 45 kA peak
current \cite{ZAGO2}. In this way, the SASE saturation power could
be increased to about 0.5 TW.

In order to generate high-power x-ray pulses we exploit undulator
tapering. Tapering consists in a slow reduction of the field
strength of the undulator in order to preserve the resonance
wavelength, while the kinetic energy of the electrons decreases due
to the FEL process. The undulator taper can be simply implemented as
discrete steps from one undulator segment to the next, by changing
the undulator gap. In this way, the output power of the SASE3
undulator could be increased from the value of 0.5 TW in the SASE
saturation regime to about 5 TW. The SASE3 undulator with 21 cells
consists of two parts. The first is composed by an uniform
undulator, the second consists of a tapered undulator. The SASE
signal is exponentially amplified passing through the first uniform
part. This is long enough, 9 cells, in order to reach saturation,
which yields about 0.5 TW power. Finally, in the second part of the
undulator the SASE output is enhanced up to 5 TW by taking advantage
of magnetic field tapering over the last 12 cells.

From all applications of XFELs for life sciences, the main
expectation and the main challenge is the determination of 3D
structures of biomolecules and their complexes from diffraction
images of single particles. Parameters of the accelerator complex
and availability of long baseline undulators at the European XFEL
offer the opportunity to build a beamline suitable for single
biomolecular imaging experiments from the very beginning of the
operation phase. In the next decade, no other infrastructure will
offer such high peak current (up to about 50 kA) and high electron
beam energy (up to about 17.5 GeV) enabling 10 TW mode of operation
in the simplest SASE regime.

\section{\label{sec:FELst} FEL Studies}

In this section we present a  feasibility study of the setup
described above with the help of the FEL code Genesis 1.3
\cite{GENE} running on a parallel machine. Results are presented for
the SASE3 FEL line of the European XFEL, based on a statistical
analysis consisting of $100$ runs. The overall beam parameters used
in the simulations are presented in Table \ref{tt1}.

\begin{table}
\caption{Parameters for the mode of operation at the European XFEL
used in this paper.}

\begin{small}\begin{tabular}{ l c c}
\hline & ~ Units &  ~ \\ \hline
Undulator period      & mm                  & 68     \\
Periods per cell      & -                   & 73   \\
Total number of cells & -                   & 21    \\
Intersection length   & m                   & 1.1   \\
Energy                & GeV                 & 17.5 \\
Charge                & nC                  & 0.25\\
\hline
\end{tabular}\end{small}
\label{tt1}
\end{table}

\begin{figure}[tb]
\includegraphics[width=0.5\textwidth]{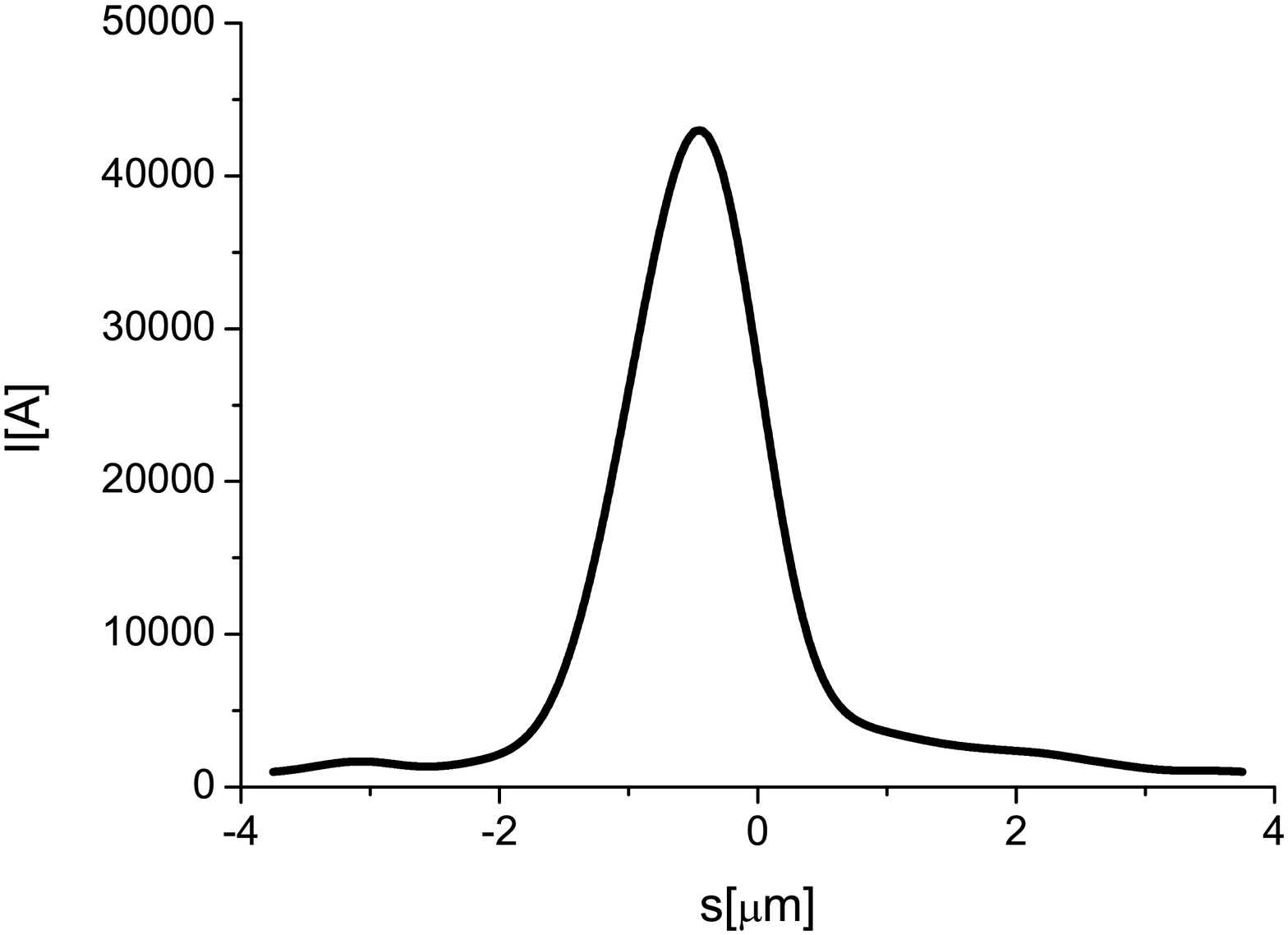}
\includegraphics[width=0.5\textwidth]{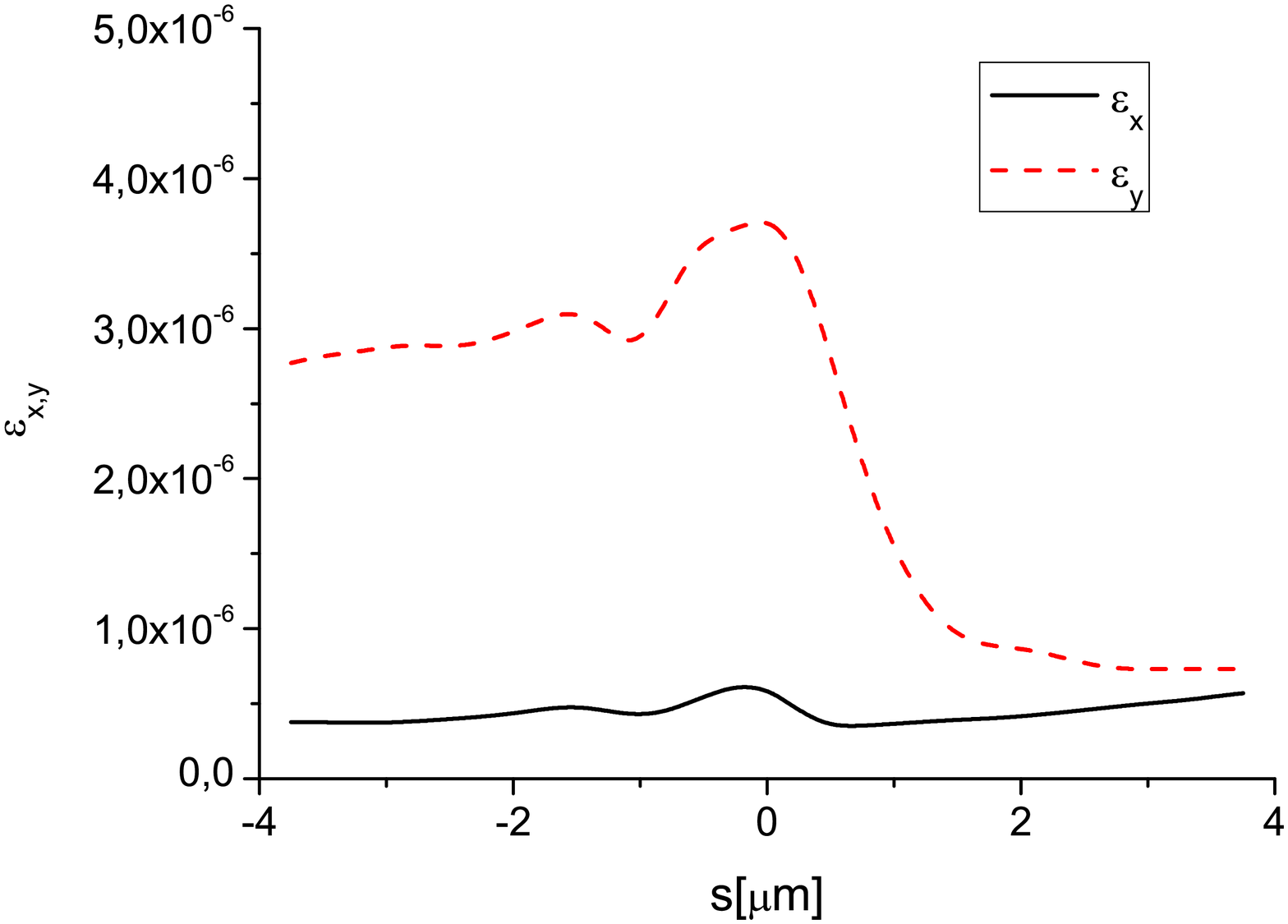}
\includegraphics[width=0.5\textwidth]{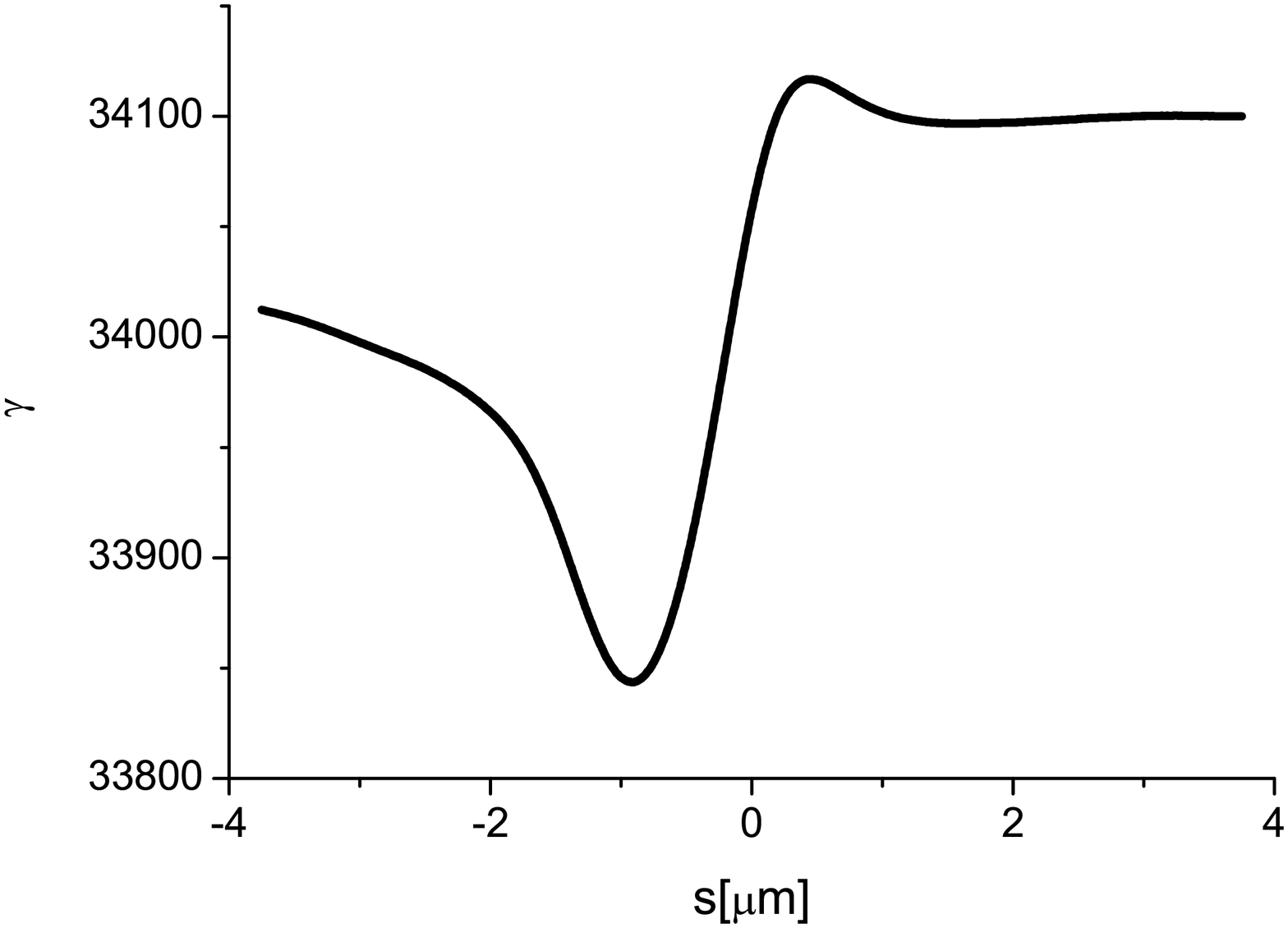}
\includegraphics[width=0.5\textwidth]{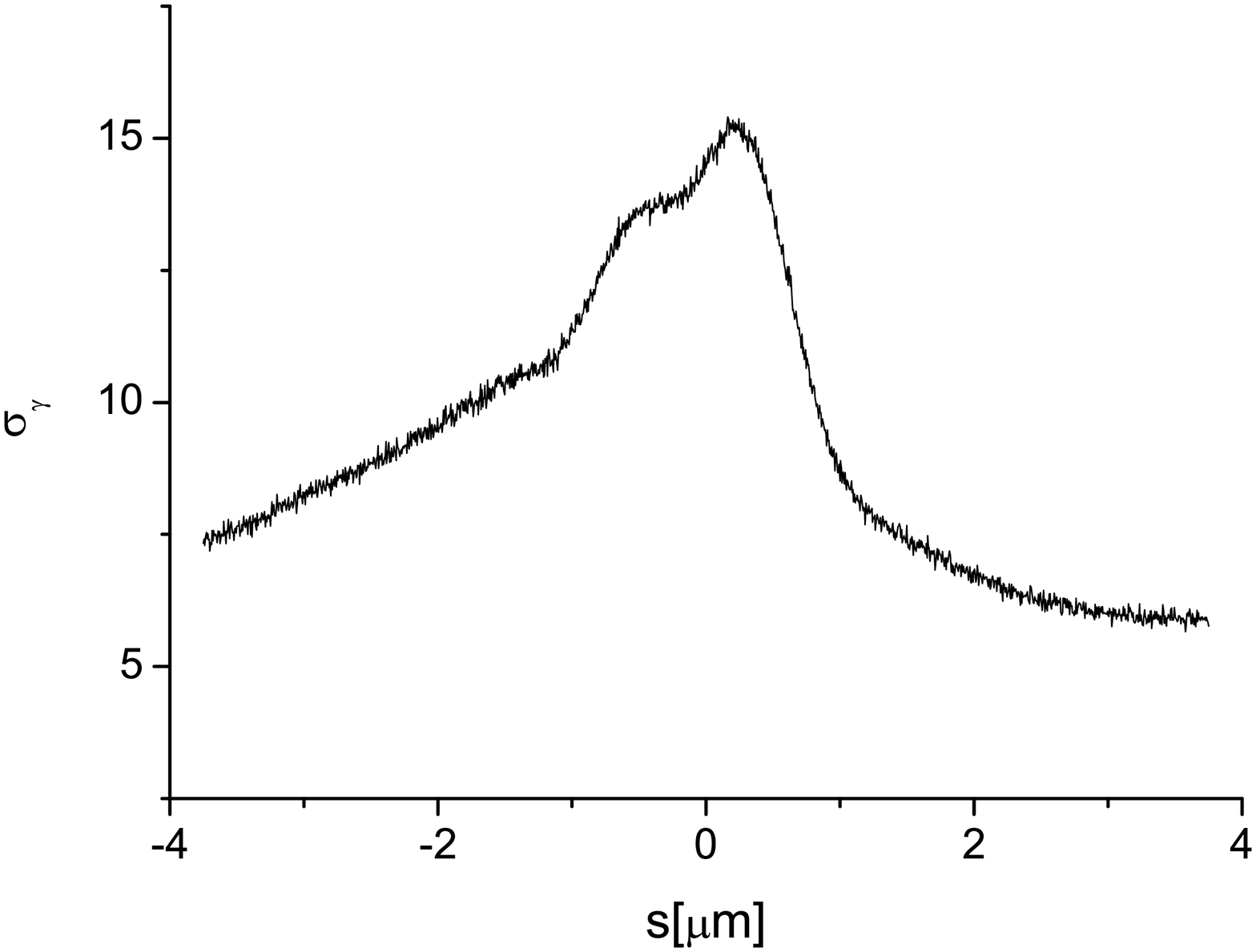}
\begin{center}
\includegraphics[width=0.5\textwidth]{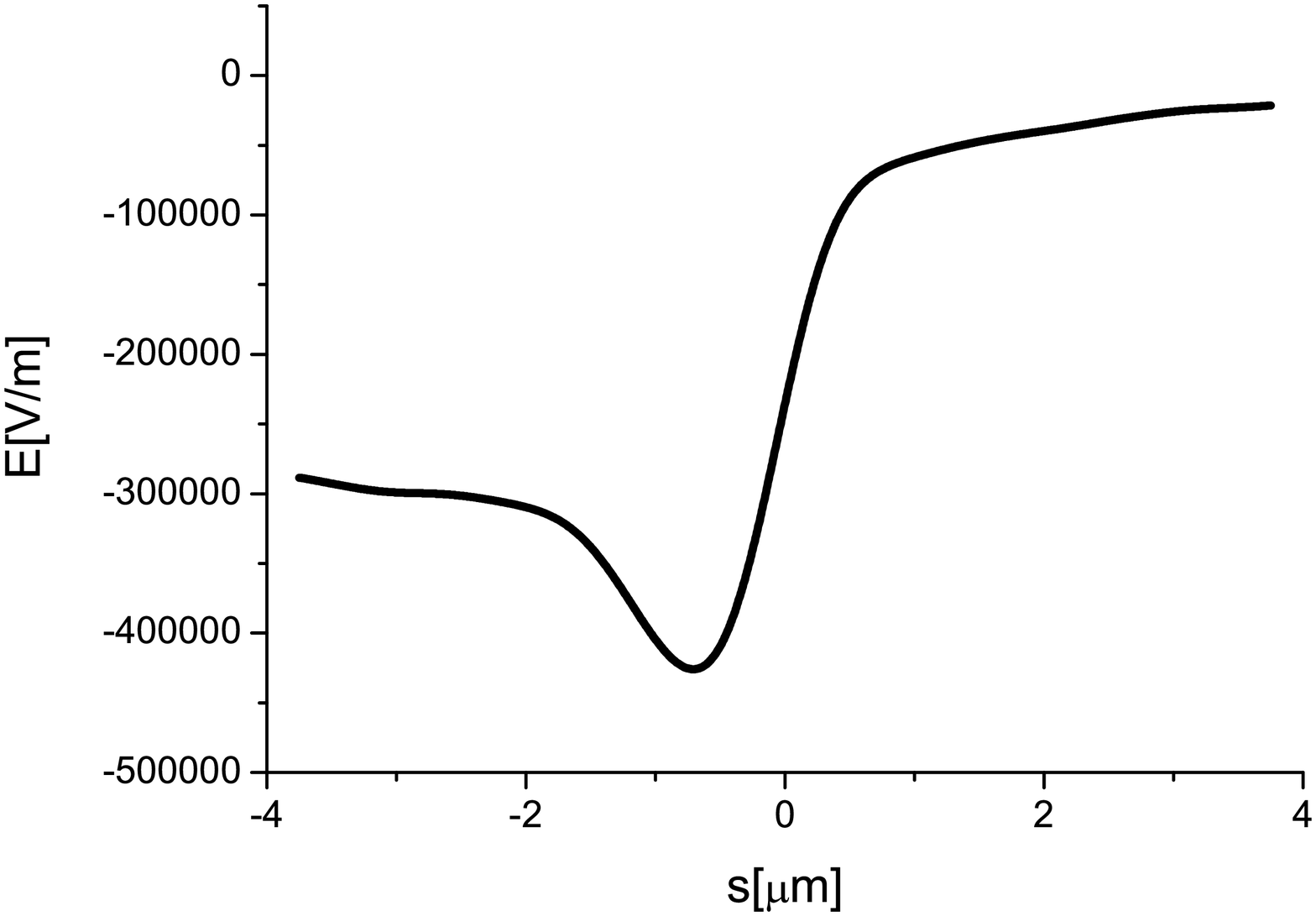}
\end{center}
\caption{Results from electron beam start-to-end simulations at the
entrance of SASE3. (First Row, Left) Current profile. (First Row,
Right) Normalized emittance as a function of the position inside the
electron beam. (Second Row, Left) Energy profile along the beam.
(Second Row, Right) Electron beam energy spread profile. (Bottom
row) Resistive wakefields in the SASE3 undulator.} \label{s2E}
\end{figure}
\begin{figure}[tb]
\begin{center}
\includegraphics[width=0.5\textwidth]{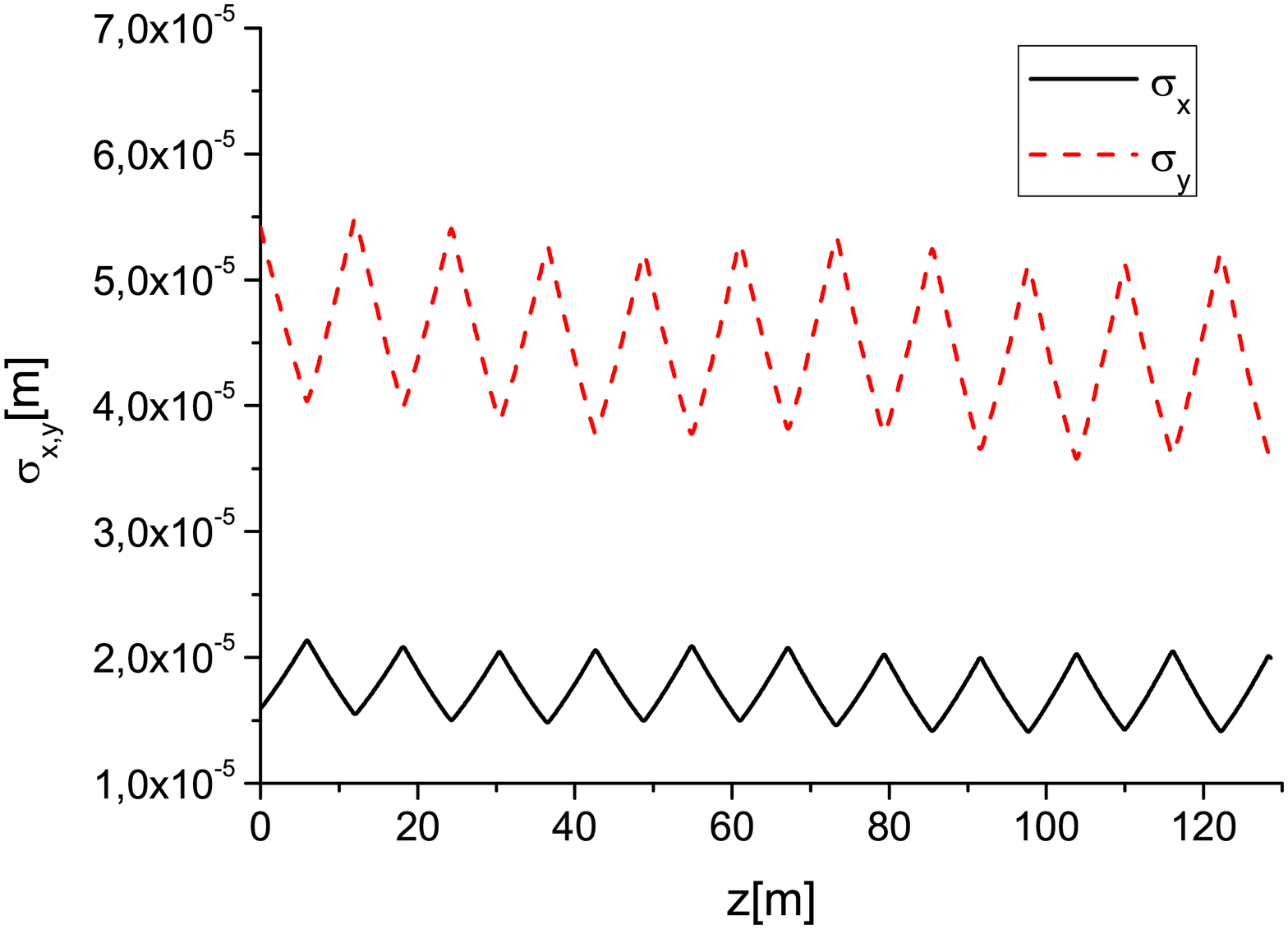}
\end{center}
\caption{Evolution of the horizontal and vertical dimensions of the
electron bunch as a function of the distance inside the SASE3
undulator. The plots refer to the longitudinal position inside the
bunch corresponding to the maximum current value.} \label{sigma}
\end{figure}
\begin{figure}
\begin{center}
\includegraphics[width=0.50\textwidth]{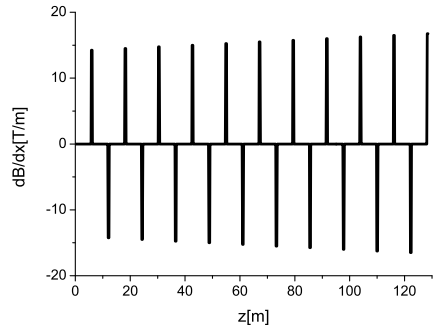}
\end{center}
\caption{Quadrupole strength along the undulator.} \label{quad}
\end{figure}
\begin{figure}
\begin{center}
\includegraphics[width=0.50\textwidth]{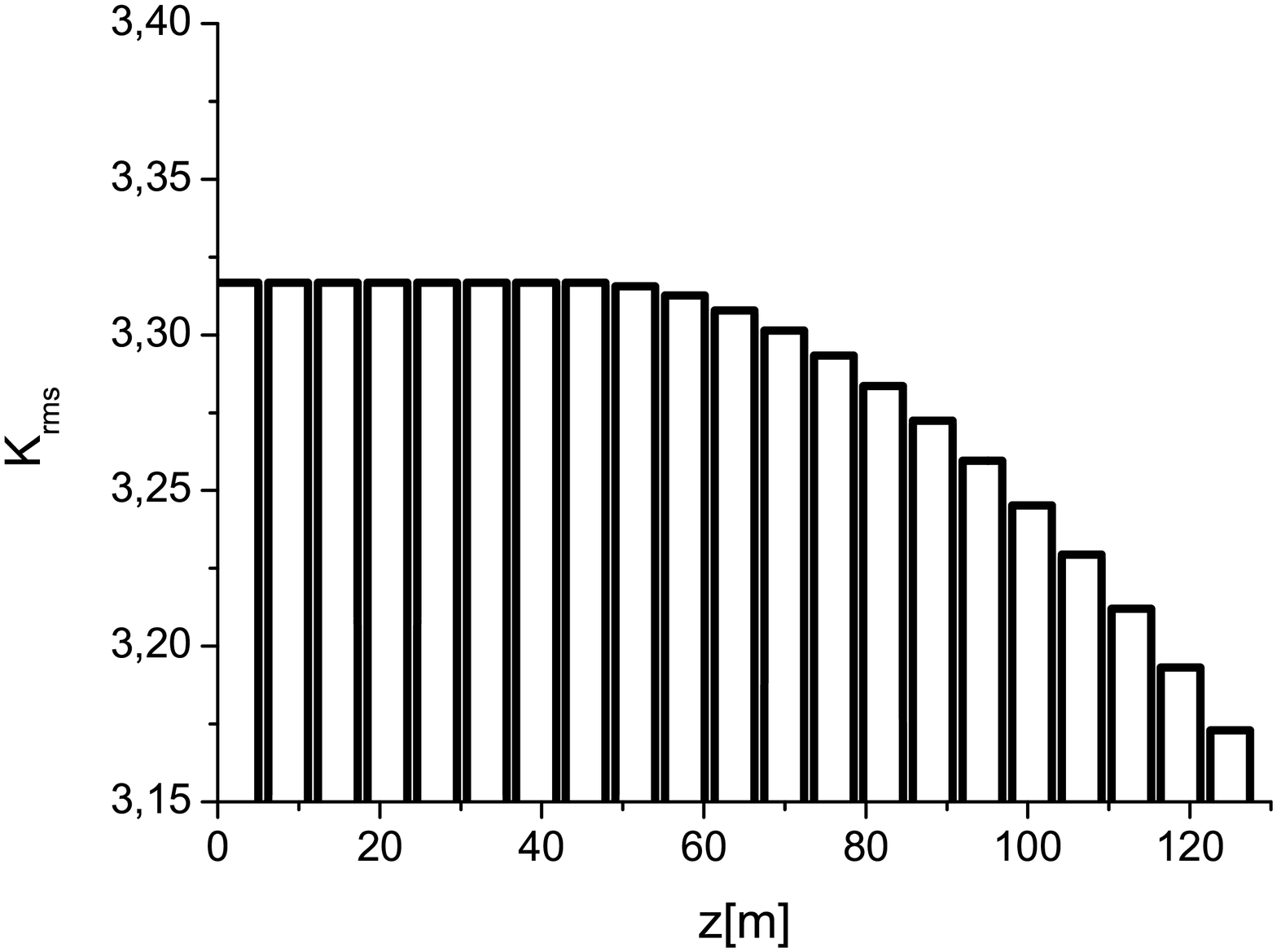}
\end{center}
\caption{Tapering law.} \label{Taplaw}
\end{figure}
The beam parameters at the entrance of the SASE3 undulator, and the
resistive wake inside the undulator are shown in Fig. \ref{s2E},
\cite{ZAGO2}. Full tracking calculations were used to find a new set
of electron bunch parameters at the entrance of baseline undulators.
The main effects influencing the electron beam acceleration and
transport, such as space charge force, rf wakefields and coherent
synchrotron radiation (CSR) effects inside magnetic compressors have
been included. Our calculations account for both wakes and quantum
fluctuations in the SASE1 undulator.

Using a bunch with larger slice emittance and energy spread, but
also higher peak current, does not necessarily complicates reaching
SASE saturation, because the increased peak current eases the
effects of the increased longitudinal velocity spread. For example,
the final normqlized slice emittance in the 45 kA case studied here
is about $4~ \mu$m, but the SASE saturation length is in the very
safe range of 9 undulator cells at photon energies around 4 keV. The
extreme working point at 45 kA peak current is very interesting,
because the radiation peak power at saturation is ten-fold increased
up to about 0.5 TW. The problem with operation at higher peak
current is that wake fields become larger and, therefore, the energy
chirp within the electron bunch becomes in its turn more and more
important. In our case of interest, the variation in the electron
energy within the bunch can be large compared to the Pierce
parameter $\rho$ ( i.e. with the slice gain-bandwidth )
\cite{ZAGO2}, but this does not result in gain reduction\footnote{In
order to incur in gain reduction, one should have a relative
variation in the electron beam energy comparable or larger than the
Pierce parameter $\rho$ within a cooperation length. }.
Specifically, simulations show that the large energy chirp along the
electron bunch only yields a large (about $1 \%$) output radiation
bandwidth.

Due to collective effects in the bunch compression system,
emittances in the horizontal and vertical directions are
significantly different. As a result, the electron beam looks highly
asymmetric in the transverse plane: in the horizontal direction
$\sigma_x \sim 20 \mu$m, while in the vertical direction $\sigma_y
\sim 50 \mu$m. The evolution of the transverse electron bunch
dimensions are plotted in Fig. \ref{sigma}. The evolution of the
transverse electron bunch dimensions is plotted in Fig. \ref{sigma},
and the correspondent quadrupole strength is shown in Fig.
\ref{quad}. The undulator is tapered according to the law in Fig.
\ref{Taplaw}. The quadrupole strength and the tapering have been
optimized to maximize the final output power.

\begin{figure}
\includegraphics[width=0.50\textwidth]{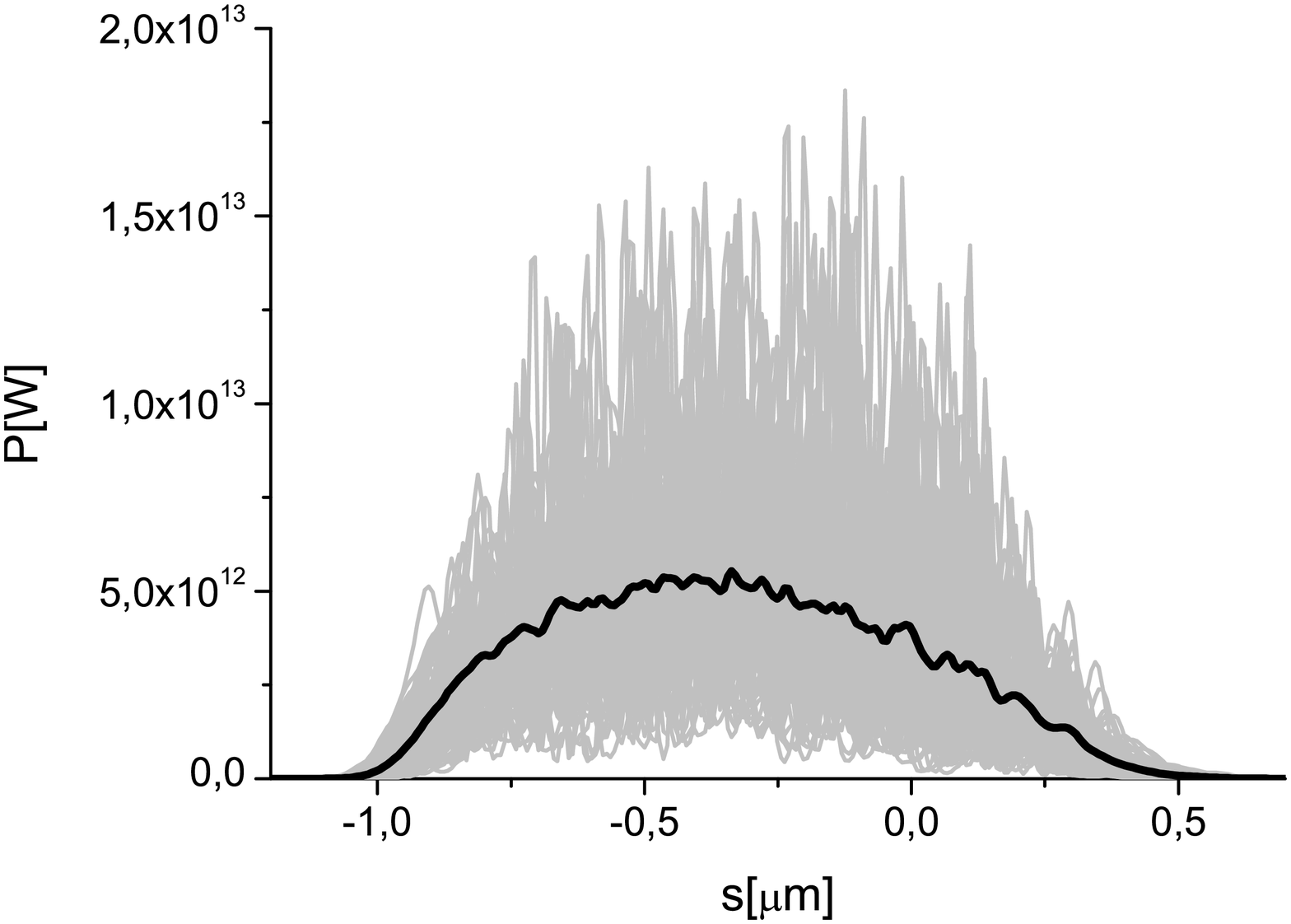}
\includegraphics[width=0.50\textwidth]{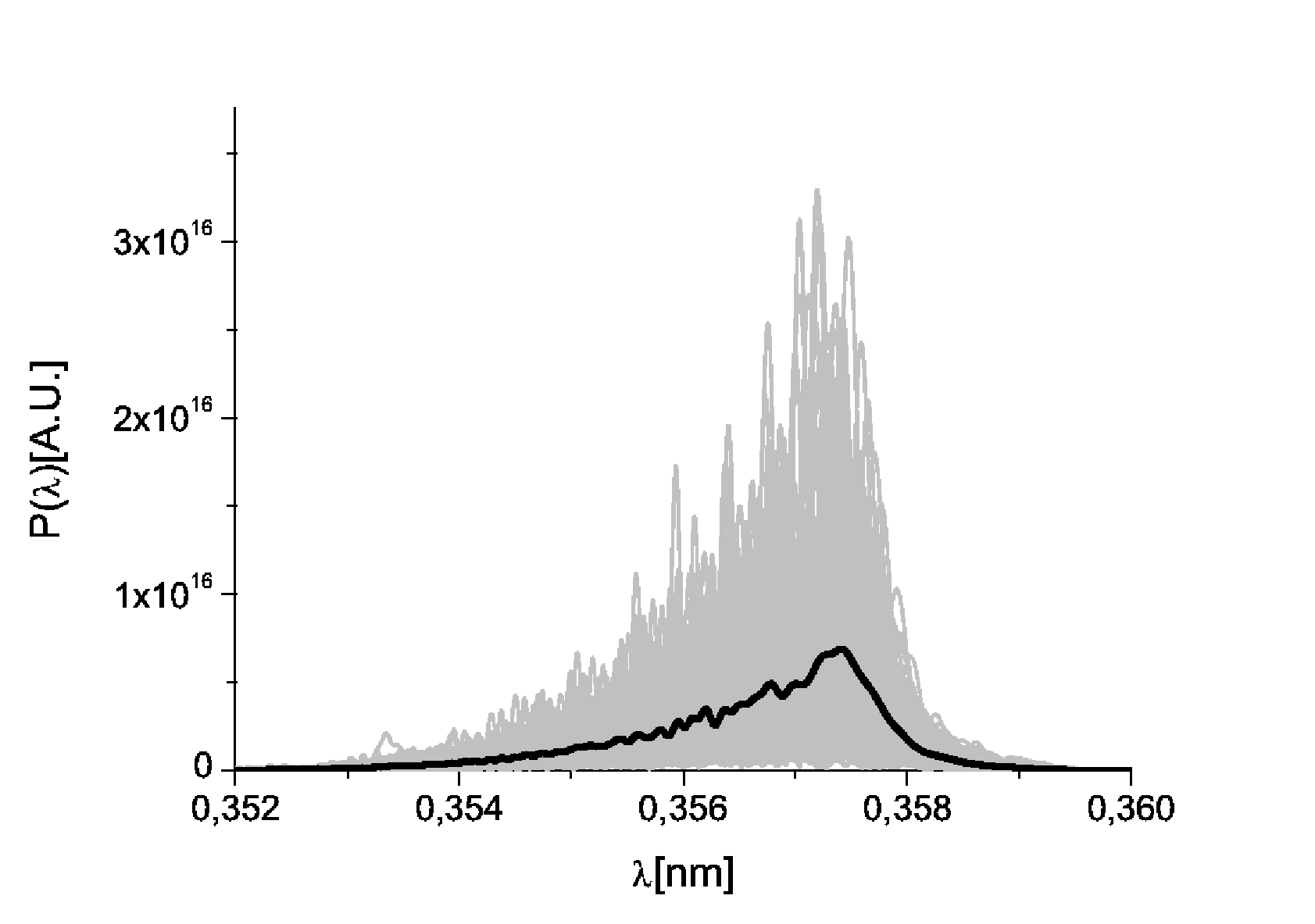}
\caption{Power and spectrum produced in the  SASE mode at saturation
with undulator tapering. Grey lines refer to single shot
realizations, the black line refers to the average over a hundred
realizations.} \label{Pout}
\end{figure}

\begin{figure}[tb]
\includegraphics[width=0.5\textwidth]{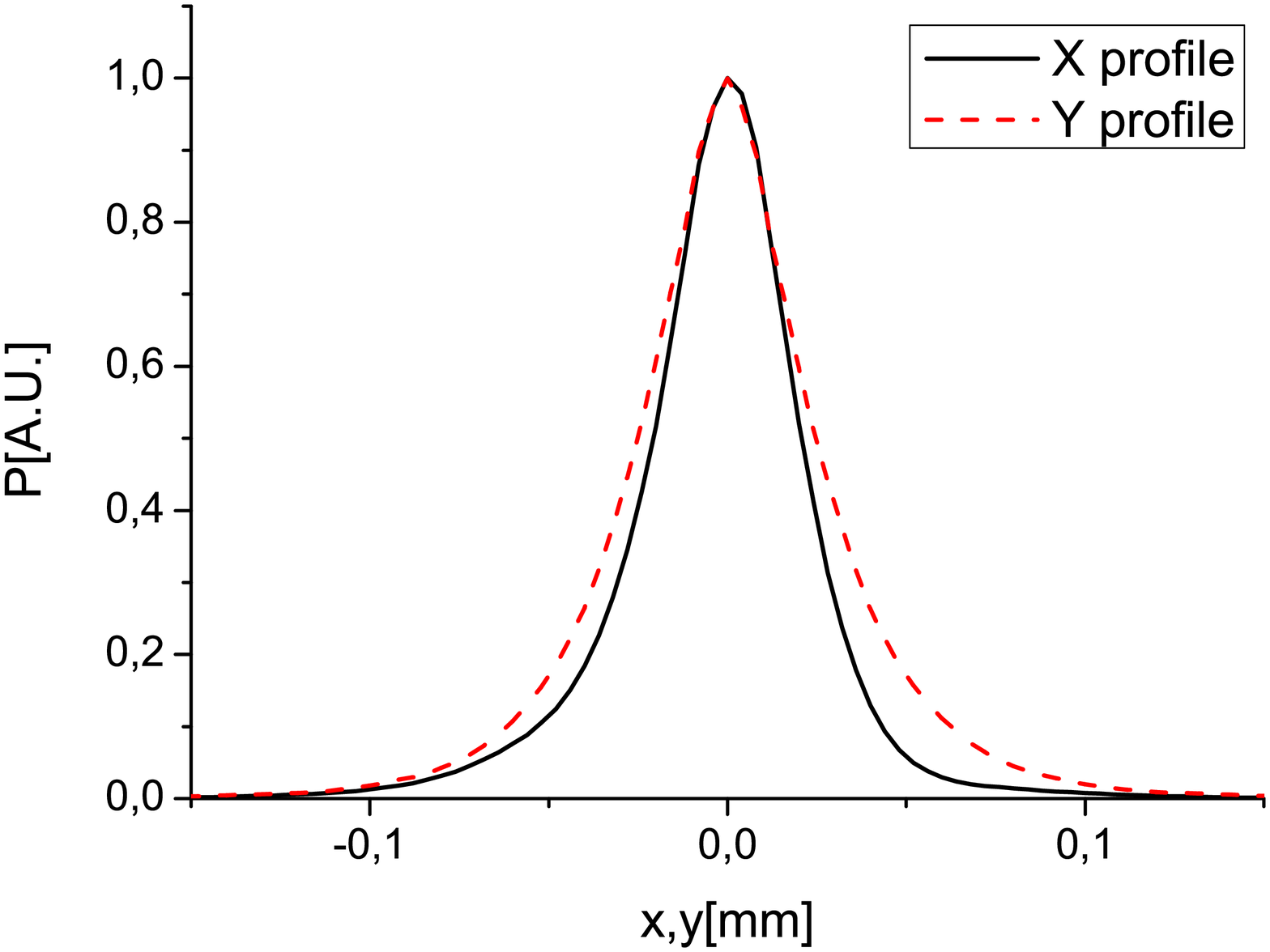}
\includegraphics[width=0.5\textwidth]{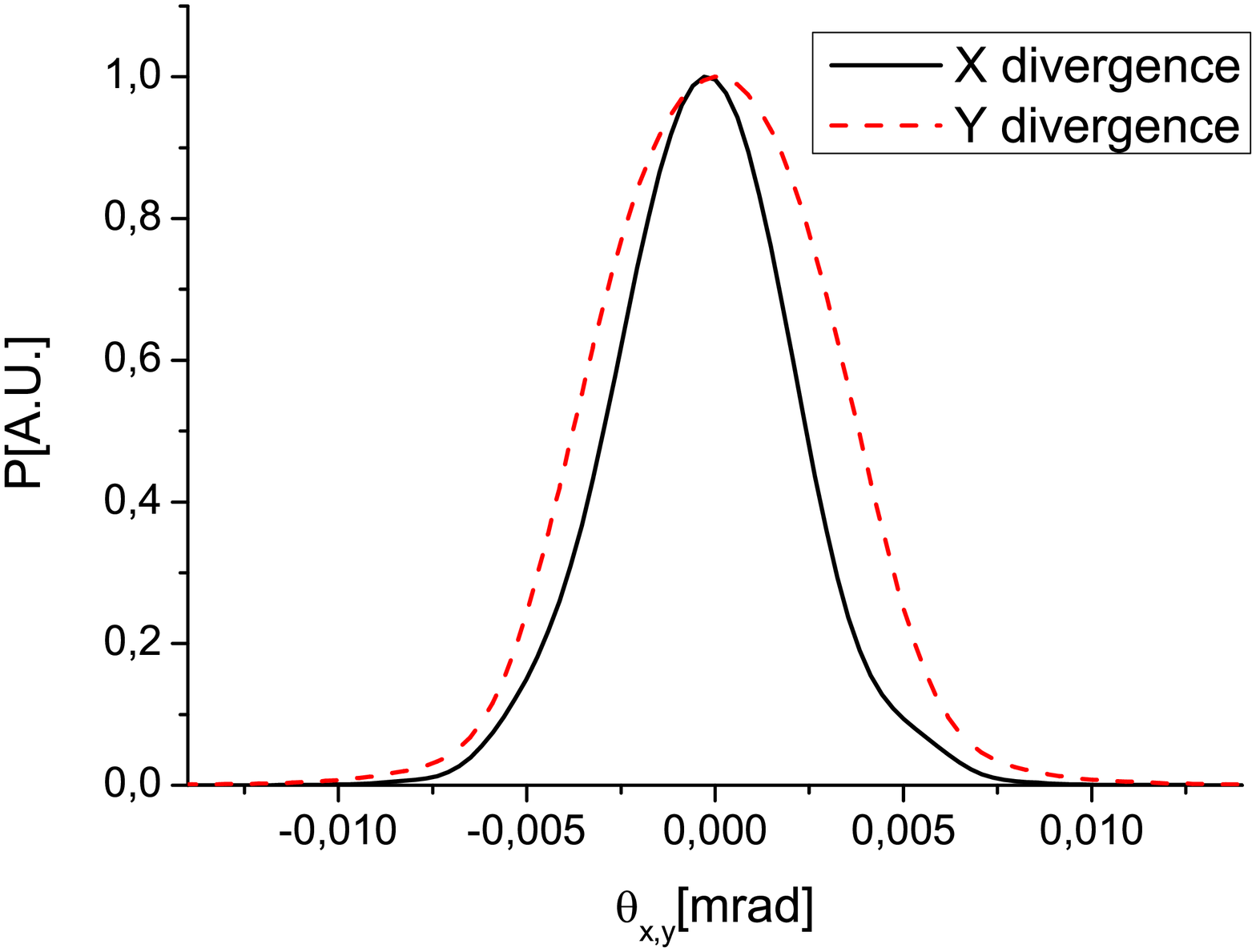}
\caption{Distribution of the radiation pulse energy per unit surface
and angular distribution of the  exit of the setup.} \label{spotTS}
\end{figure}

\begin{figure}
\includegraphics[width=0.50\textwidth]{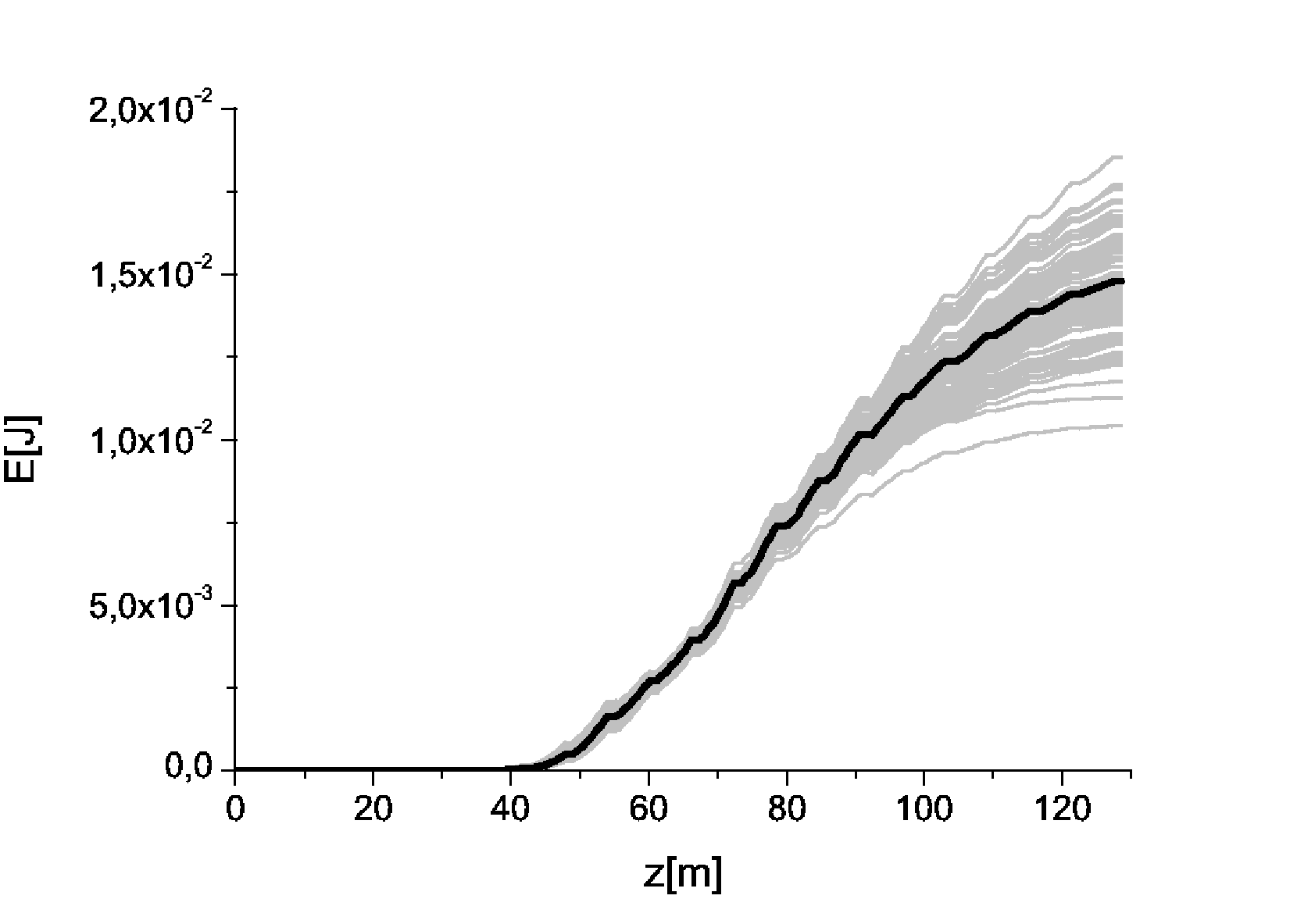}
\includegraphics[width=0.50\textwidth]{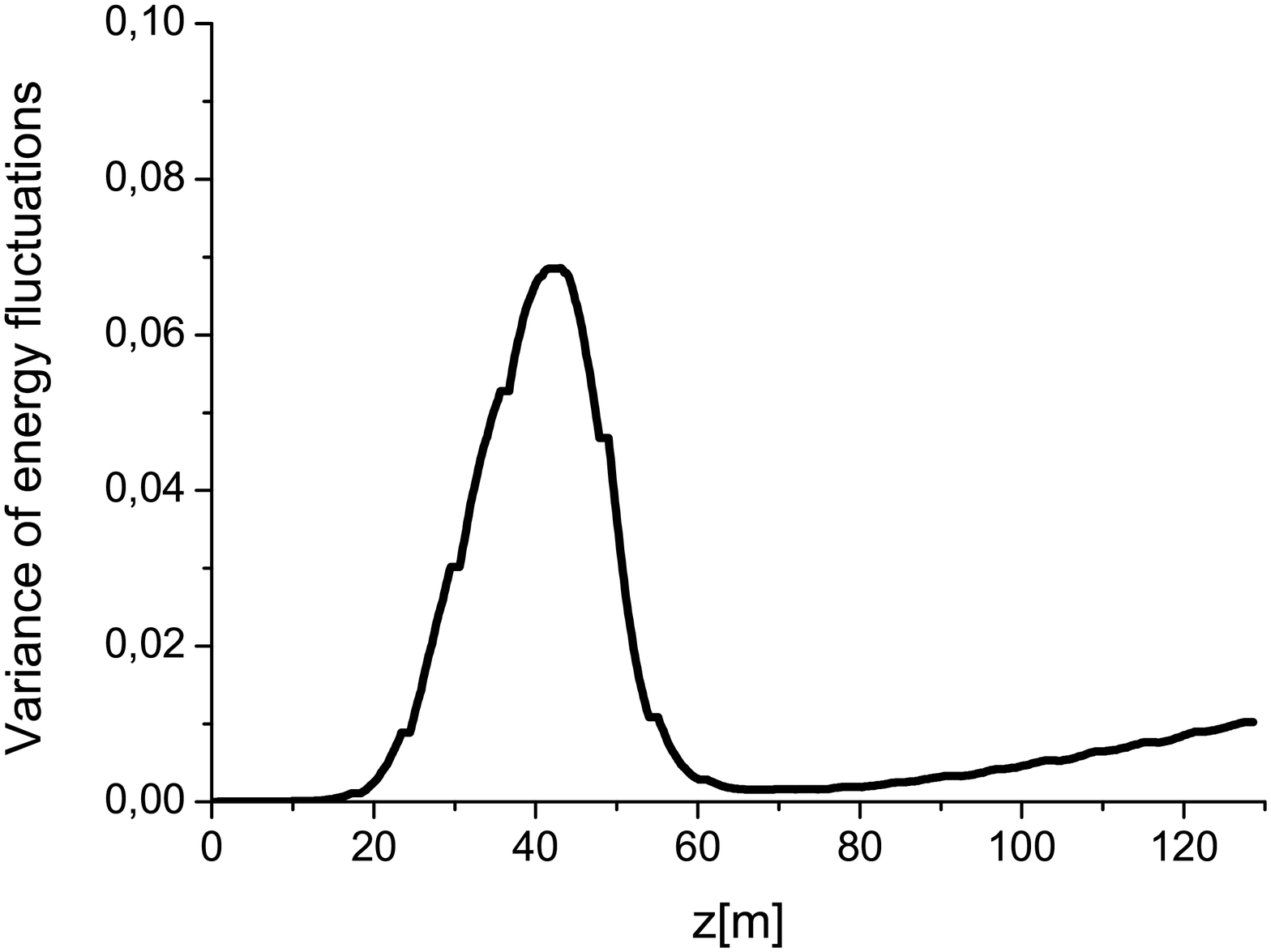}
\caption{Evolution of the output energy in the photon pulse and of
the variance of the energy fluctuation as a function of the distance
inside the output undulator, with tapering. Grey lines refer to
single shot realizations, the black line refers to the average over
a hundred realizations.} \label{EnvaroutS}
\end{figure}
The output characteristics, in terms of power and spectrum, are
plotted in Fig. \ref{Pout}. Inspection of the plots shows that one
can reach $5$ TW pulses with a bandwidth of about $1 \%$. Fig.
\ref{spotTS} shows the distribution of the radiation pulse energy
per unit surface and angular distribution of the  exit of the setup.
Finally, in Fig. \ref{EnvaroutS} we plot the evolution of the output
energy in the photon pulse and of the variance of the energy
fluctuation as a function of the distance inside the output
undulator.

\section{\label{sec:concl} Conclusions}

The nominal design parameters for the European XFEL for a 0.25 nC
electron bunch, which allow for SASE saturation with 0.4 $\mu$m
normalized slice emittance and 5 kA peak current are described in
\cite{ZAGO1}. In this article we note that the European XFEL
accelerator complex is flexible enough to be reconfigured for much
higher bunch peak-current. In this case, the new beam parameters are
simply set in the control room, and do not require hardware
modifications in the tunnel. This flexibility is demonstrated by
studying the new acceleration and compression parameters required
over a wide range of a peak current values well beyond the nominal 5
kA \cite{ZAGO2}. For each case, full tracking calculations were used
to find a new set of electron bunch parameters at the entrance of
baseline undulators. The main effects influencing the electron beam
acceleration and transport, such as space charge force, rf
wakefields and coherent synchrotron radiation (CSR) effects inside
magnetic compressors have been included. Using a bunch with larger
slice emittance and energy spread, but also higher peak current,
does not necessarily complicates reaching SASE saturation, because
the increased peak current eases the effects of the increased
longitudinal velocity spread. For example, the final slice
normalized emittance in the 45 kA case studied here is about $4
~\mu$m, but the SASE saturation length is in very safe range of 9
undulator cells at photon energies around 4 keV. The extreme working
point at 45 kA peak current is very interesting, because the
radiation peak power at saturation is ten-fold increased up to about
0.5 TW. The problem with operation at higher peak current is that
wake fields become larger and, therefore, the energy chirp within
the electron bunch becomes in its turn more and more important. In
our case of interest, the variation in the electron energy within
bunch can be large compared to the Pierce parameter $\rho$ ( i.e.
with the slice gain-bandwidth ) \cite{ZAGO2}, but this does not
result in gain reduction. Specifically, simulations show that the
large  energy chirp along the electron bunch only yields a large
(about $1 \%$) output radiation bandwidth.

For certain experiments, a high degree of monochromacity is not
needed. An important class of experiments when a large radiation
bandwidth is not detrimental is single biomolecular imaging. In such
instances, photon fluence rather than spectral photon density is a
premium. One obvious way to enhance the SASE efficiency is by
properly configuring undulators with variable gap. It has been shown
that this approach allows one to increase the peak power to 5 TW by
taking advantage of an undulator magnetic-field taper over the
baseline SASE3 undulator. Such unique feature will be available only
at the European XFEL for a long time. In fact, the European XFEL
uses a super-conducting L-band linac to accelerate beams which can
be compressed up to extremely  high peak current (about 50 kA), with
still a reasonable electron bunch quality. Moreover, high electron
beam energy (up to about 17.5 GeV) and long undulators with high K
values imply a high peak output-power.

\section{Acknowledgements}

We are grateful to Massimo Altarelli, Reinhard Brinkmann, Henry
Chapman, Janos Hajdu, Viktor Lamzin, Serguei Molodtsov and Edgar
Weckert for their support and their interest during the compilation
of this work.


\begin{thebibliography}{99}


\bibitem{HAJD} J. Hajdu, Curr. Opin. Struct. Biol. 10, 569 (2000)

\bibitem{NEUT} R. Neutze et al., Nature 406, 752 (2000)

\bibitem{CHA2} H. Chapman et al., Nat. Phys. 2, p. 839 (2006).

\bibitem{CHAP} K. J. Gaffney and H. N. Chapman, Science 316, 1444 (2007)

\bibitem{SEIB} M. M. Seibert et al., Nature 470 (7332) 78-81 (2011)

\bibitem{BERG} S. Baradaran et al., LCLS-II New Instruments Workshops Report, SLAC-R-993
(2012), see Section 4.3.2. by H. Chapman et al., and Section 4.3.3.
by F. R. N. C. Maia et al.


\bibitem{SELF0} S. Serkez et al., Proposal for a scheme to generate 10 TW-level femtosecond x-ray pulses for imaging single protein molecules at the European
XFEL, DESY 13-101, http://arxiv.org/abs/1306.0804 (2013).

\bibitem{TSCH} Th. Tschentscher, "Layout of the x-Ray Systems at the European
XFEL", Technical Report 10.3204/XFEL.EU/TR-2011-001 (2011).




\bibitem{SELF} J. Feldhaus et al., Optics. Comm. 140, 341 (1997).

\bibitem{SXFE} E. Saldin, E. Schneidmiller,  Yu. Shvyd'ko and M.
Yurkov, NIM A 475 357 (2001).

\bibitem{SOPT} E. Saldin, E. Schneidmiller and M. Yurkov, NIM A 445
178 (2000).


\bibitem{STTF} R. Treusch, W. Brefeld, J. Feldhaus and U Hahn,  Ann. report 2001
"The seeding project for the FEL in TTF phase II" (2001).

\bibitem{SCOM} A. Marinelli et al., Comparison of HGHG and Self Seeded Scheme for the Production of Narrow Bandwidth FEL
Radiation, Proceedings of FEL 2008, MOPPH009, Gyeongju (2008).

\bibitem{OURL} G. Geloni, V. Kocharyan and E.~Saldin, "Scheme for generation of highly monochromatic X-rays from a baseline
XFEL  undulator", DESY 10-033 (2010).


\bibitem{HUAN} Y. Ding, Z. Huang and R. Ruth, Phys.Rev.ST Accel.Beams, vol. 13, p. 060703 (2010).

\bibitem{OURX} G. Geloni, V. Kocharyan and E.~Saldin, "A simple method for controlling the line width of SASE X-ray FELs",
DESY 10-053 (2010).

\bibitem{OURY2} G. Geloni, V. Kocharyan and E.~Saldin, "A Cascade self-seeding scheme with wake monochromator for narrow-bandwidth X-ray FELs", DESY 10-080 (2010).


\bibitem{OURY4} Geloni, G., Kocharyan, V., and Saldin, E., "Cost-effective way to enhance the capabilities of the LCLS baseline", DESY 10-133 (2010).




\bibitem{WU} J. Wu et al., "Staged self-seeding scheme for narrow
bandwidth , ultra-short X-ray harmonic generation free electron
laser at LCLS", proceedings of 2010 FEL conference, Malmo, Sweden,
(2010).


\bibitem{OURY6} Geloni, G., Kocharyan V., and Saldin, E., "Generation of doublet spectral lines at self-seeded X-ray FELs", DESY 10-199
(2010), and Optics Communications,  284, 13, 3348 (2011)


\bibitem{OURY5} Geloni, G.,  Kocharyan, V.,  and Saldin, E., "Production of transform-limited X-ray pulses through
self-seeding at the European X-ray FEL", DESY 11-165 (2011).


\bibitem{OURY5b} Geloni, G., Kocharyan V., and Saldin, E., "A novel Self-seeding scheme for hard X-ray FELs", Journal of Modern
Optics, vol. 58, issue 16, pp. 1391-1403,
DOI:10.1080/09500340.2011.586473 (2011)




\bibitem{WUFEL2} J. Wu et al., Simulation of the Hard X-ray
Self-seeding FEL at LCLS, MOPB09,  FEL 2011 Conference proceedings,
Shanghai, China (2011).

\bibitem{AMAN} J. Amann et al., Nature Photonics, DOI:
10.1038/NPHOTON.2012.180 (2012).

\bibitem{SHVI} R. R. Lindberg and Yu.V. Shvyd'ko, Phys. Rev. ST Accel. Beams 15,
100702 (2012).

\bibitem{FENG3} Y. Feng et al., "System design for self-seeding the
LCLS at soft X-ray energies", Proceedings of the 24th International
FEL Conference, Nara, Japan (2012).

\bibitem{GRAT} S. Serkez, G. Geloni, V. Kocharyan and
E. Saldin, "Grating monochromator for soft X-ray self-seeding the
European XFEL", DESY 13-040, http://arxiv.org/abs/1303.1392 (2013).


\bibitem{ASYM} G. Geloni, V. Kocharyan, E. Saldin, S. Serkez and M.
Tolkiehn, "Wake monochromator in asymmetric and symmetric Bragg and
Laue geometry for self-seeding the European XFEL", DESY 13-013
(2013).


\bibitem{EMM1} P. Emma, K. Bane, M. Cornacchia, Z. Huang, H. Schlarb, G. Stupakov, and D. Walz, PRL 92, 074801-1
(2004).

\bibitem{EMM2} P. Emma, M. Borland and Z. Huang, Proceedings of the FEL Conference
2004, TUBIS01, p.333 (2004).

\bibitem{DING} Y. Ding , et al., PRL 109, 254802-1 (2012).

\bibitem{TAP1} A. Lin and J.M. Dawson, Phys. Rev. Lett. 42 2172
(1986).

\bibitem{TAP2} P. Sprangle, C.M. Tang and W.M. Manheimer, Phys. Rev. Lett. 43 1932
(1979).

\bibitem{TAP3} N.M. Kroll, P. Morton and M.N. Rosenbluth, IEEE J. Quantum Electron., QE-17, 1436
(1981).

\bibitem{TAP4} T.J. Orzechovski et al., Phys. Rev. Lett. 57, 2172
(1986).

\bibitem{FAWL} W. Fawley et al., NIM A 483 p 537 (2002).

\bibitem{CORN} M. Cornacchia et al.,  J. Synchrotron rad. 11,
227-238 (2004).

\bibitem{WANG} X. Wang et al., PRL 103, 154801 (2009).


\bibitem{OURY3} G. Geloni, V. Kocharyan and E.~Saldin, "Scheme for generation of fully coherent, TW power level hard x-ray pulses from baseline undulators at the European XFEL", DESY 10-108 (2010).




\bibitem{WUFEL1} W.M. Fawley et al., Toward TW-level LCLS radiation
pulses, TUOA4, FEL 2011 Conference proceedings, Shanghai, China
(2011).

\bibitem{LAST} Y. Jiao et al. Phys. Rev. ST Accel. Beams 15, 050704
(2012).




\bibitem{DOHL} T. Limberg et al., "Optimized bunch compressor system for
the European XFEL", Proceedings of 2005 Particle Accelerator
Conference, Knoxville, Tennessee, 1236,
http://accelconf.web.cern.ch/AccelConf/P05/PAPERS/RPPT011.PDF (2005)


\bibitem{ZAGO2} I. Zagorodnov, "Compression scenarios for the European XFEL",
http://www.desy.de/fel-beam/data/talks/files/Zagorodnov$\_$ACC2012$\_$ready$\_$new.pptx,
(2012).

\bibitem{GENE} S. Reiche et al., Nucl. Instr. and Meth. A 429, 243 (1999).

\bibitem{ZAGO1} I. Zagorodnov, "Beam Dynamics Simulations for XFEL",
http://www.desy.de/xfel-beam/s2e (2011).



\end{thebibliography}
\end{document}